\documentstyle [12pt,epsf] {article}

\catcode`@=12
\newcommand{\be}{\begin{equation}}
\newcommand{\ee}{\end{equation}}
\newcommand{\ber}{\begin{eqnarray}}
\newcommand{\eer}{\end{eqnarray}}
\newcommand{\bers}{\begin{eqnarray*}}
\newcommand{\eers}{\end{eqnarray*}}
\newcommand{\bt}{\begin{itemize}}
\newcommand{\et}{\end{itemize}}
\begin{document}
\vspace{0.5in}
\oddsidemargin -.375in
\newcount\sectionnumber
\sectionnumber=0
\def\bra#1{\left\langle #1\right|}
\def\ket#1{\left| #1\right\rangle}
\def\be{\begin{equation}}
\def\ee{\end{equation}}
\thispagestyle{empty}
\begin{flushright} UdeM-GPP-TH-02-94 \\
UTPT-02-01,
TAUP 2696-02,
WIS/07/02-JAN-DPP\\
February 2002\
\end{flushright}
\vspace {.5in}
\begin{center}
{\Large\bf Charmless B Decays to Final States with
Radially Excited  Vector Mesons   \\}
\vspace{.5in}
{{\bf Alakabha Datta{\footnote{email: datta@lps.umontreal.ca}}
${}^{a}$}, {\bf Harry J. Lipkin {\footnote{email:
harry.lipkin@weizmann.ac.il}}
${}^{b}$} and
{\bf Patrick J. O'Donnell {\footnote{email:
pat@medb.physics.utoronto.ca}}${}^{c}$}
\\}
\vspace{.1in}
${}^{a}$
{\it Laboratoire Ren\'e J.-A. L\'evesque, Universit\'e de
Montr\'eal,} \\
{\it C.P. 6128, succ.\ centre-ville, Montr\'eal, QC, Canada H3C 3J7} \\
${}^{b)}$ {\it
Department of Particle Physics,\\
Weizmann Institute,\\
Rehovot 76100, Israel \\and\\
School of Physics and Astronomy, \\
Tel-Aviv University,\\
Tel-Aviv 69978, Israel \\
${}^{c}$ {\it Department of Physics and Astronomy,\\
University of Toronto, Toronto, Canada.}\\
}
\end{center}

\begin{abstract}
We consider the weak decays of a B meson to final states that contain
a S-wave radially excited vector meson. We  consider vector-pseudoscalar
final states and calculate ratios
 of the type $B \to \rho^{\prime} \pi/B \to \rho \pi$,
$B \to \omega^{\prime} \pi/B \to \omega \pi$ and
$B \to \phi^{\prime} \pi/B \to \phi \pi$
where $\rho^{\prime}$, $\omega^{\prime}$ and $\phi^{\prime}$
are higher $\rho$, $\omega$
and $\phi$ S-wave radial excitations. We find
such decays to have  larger or similar branching ratios
compared
to decays where
the final state $\rho$, $\omega$ and $\phi$ are in the ground state.
We also study the effect of radial mixing in the vector system 
generated from hyperfine interaction
and the annihilation
term. 

\end{abstract}

\newpage \pagestyle{plain}

The new data accumulating from B factories and other accelerators will
include transitions to many new final states which have not been
previously studied in detail; e.g. radially excited meson states. Many
decays involve a transition from a low momentum spectator quark to a high
momentum relativistic meson. The form factors for such transitions are
expected to be sensitive to the high momentum components of the final
meson wave function, and therefore to favor radially excited states. 
 The data on B decays to these
states will thus provide important new information, particularly for the form
factors to the radially excited states and probe
the high-momentum tails of their wave functions.

\begin{figure}[htb]
   \centerline{\epsfysize 2.2 truein \epsfbox{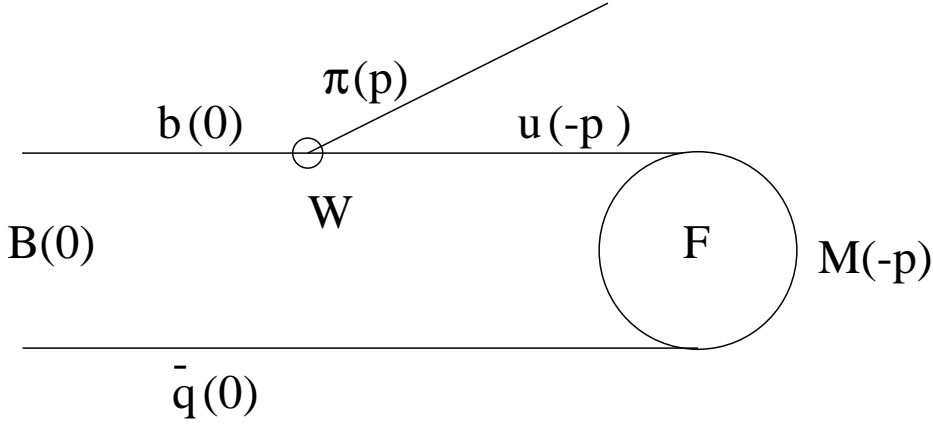}}
   \caption{Factorization for the decay $B \to M \pi$. }
\label{fac}
   \end{figure}

In this paper we calculate predictions for the ratios
\ber
R_{\rho^+} & = & BR({\bar{B}^0} \rightarrow \pi^- \rho^{+ \prime} )/
BR({\bar{B}^0} \rightarrow \pi^- \rho^{+} )
\label{1}
\eer
\ber
R_{\rho^0} & = & BR({\bar{B}^-} \rightarrow \pi^- \rho^{0 \prime} )/
BR({\bar{B}^-} \rightarrow \pi^- \rho^{0} )
\label{2}
\eer
\ber
R_{\omega} & = & BR(B^- \rightarrow \pi^- \omega' )/BR(B^- \rightarrow \pi^- 
\omega)
\label{3}
\eer
\ber
R_{\phi} & = & BR(B_s \rightarrow \pi^0 \phi' )/BR(B_s \rightarrow \pi^0 
\phi)
\label{4}
\eer
where $ \rho^{\prime}, \omega^{\prime} $ and $\phi^{\prime}$
are the radially excited
states.
Most studies of two body nonleptonic $B$ decays have concentrated on
processes of the type $B \to M_1 M_2$ where both $M_1$ and $M_2$
are mesons in the ground state configuration. Nonleptonic decays, where
one of the mesons in the final state containing the spectator quark is a 
radially excited 
state,
are expected to have  larger or similar branching ratios
compared
to decays where
the final state contains the same meson in the ground state. This is
easily seen in a simple
 model in which
$B \rightarrow \pi M$ and M is a simple flavor eigenstate with no flavor
mixing beyond isospin.
 We follow the inactive spectator approach used \cite{Datta} to
treat B decays to charmonium in which the spectator quark does not
participate in a flavor-changing interaction and later combines with a
light antiquark to make the final light meson
as shown in Fig.~\ref{fac}.  The decay amplitudes are
then described as the product of a b-quark decay amplitude and a
hadronization function describing the combination of a quark-antiquark
pair to make the final meson.
Neglecting the relative Fermi momentum of the b quark and the spectator quark
in the $B$ meson,
   the quark transition  
for the processes in Eq. (1-3) is 
\begin{equation}
b \rightarrow \pi^-(\vec p) u(-\vec p)
\label{bdecay1}
\end{equation}
where the b quark is at rest and $\vec p$ denotes 
the final momentum of the $\pi^-$.
For the process in Eq.\ref{4}, the quark transition is essentially similar to
the one above
\begin{equation}
b \rightarrow \pi^0(\vec p) s(-\vec p)
\label{bdecay2}
\end{equation}
where now $\vec p$ denotes the final momentum of the $\pi^0$.

Concentrating on the processes in Eq. (1-3)  
the transition matrix for the full decay has the form\cite{Datta}
\begin{equation}
\bra {\pi^-(\vec p) M(-\vec p)}T\ket {B} =
\bra {M(-\vec p)}F \ket{u(-\vec p) \bar q(0)}\cdot
\bra {\pi^-(\vec p) u(-\vec p) ) }W \ket {b}
\label{had}
\end{equation}
where $T$ denotes the transition matrix for the hadronic decay which
factors, as shown in Fig.~\ref{fac},
into a weak matrix element at the quark level denoted by $W$ and a
recombination
matrix element denoted by $F$. This latter matrix element  describes the
transition of a quark with
momentum
$- \vec p$ and an antiquark with zero momentum to make a meson with momentum
$- \vec p$.
  Radial excitations could be favoured over  ground
states, if the final momentum $ \vec p$ is large, since the radial 
excitations
are expected to have higher kinetic
energies.

An alternative but equivalent way of understanding this effect is to note 
that
for a heavy $b$ quark one can write
\cite{Beneke}
\begin{equation}
\bra {\pi^-(\vec p) M(-\vec p)}T\ket {B} =
\bra {M(-\vec p)}J_{1\mu} \ket{B}\cdot
\bra {\pi^-(\vec p)}J^{\mu}_{2} \ket {0}\
\end{equation}
where $J_{1,2}$ are currents that occur in $W=J_1 \times J_2$. The 
transition
matrix element for the hadronic decay can then be written in terms of
$B \to M$
form factors
and the pion decay constant. The form factors can be expressed as
overlap integrals of the $B$ and
the $M$ meson wavefunctions.
When $M$ is a light meson, with a mass much smaller than the $B$ meson,
the overlap
integrals get contributions mainly from the high momentum components
of the meson wavefunctions. It is now clear that for a radially excited 
meson
$M^{\prime}$, which has more higher momentum components, the overlap integrals will
be enhanced compared to the overlap integral with a ground state meson $M$.
Consequently the $B \to M^{\prime}$ form factors are likely to be enhanced relative
to the
$B \to M$ form factors which would then translate into higher branching 
ratio for
$B \to M^{\prime} \pi $ relative to $B \to M \pi$.

Our discussion above assumed the physical states to be pure radial 
excitations.
However, additional interactions can mix the various 
radial excited components.
For instance hyperfine interactions can mix radial
excitations with the same flavor structure and so in general, in
the
$\rho$, $\omega$ and $\phi$ systems, the various physical states will
be admixtures of radial excitations \cite{Cohen:1979ge,Frank:1984bj}.
Flavor mixing in the vector system
is known to be small but is important
in the pseudoscalar sector. We will consider the pseudoscalar case
in a different publication\cite{newpaper}.
To make quantitative predictions we
use constituent quark wave functions with several potentials to test the
dependence of the results on the confining potential. We shall see that
the effects of the potential dependence and mixing are small so that the
results are reasonably robust and are not seriously dependent on the
fine details of the model.

Even though our discussion has so far only included vector-pseudoscalar 
final states we can also consider vector-vector final states such as 
$B \to \rho \rho$ or $B \to J/\psi K^*$. However vector- vector final states are
complicated since different partial waves are present. Our purpose here is to demonstrate the effects of radial mixing in the simpler physical system
of the vector-pseudoscalar final state. 
If the effects of 
radial enhancements are observed in the  vector-pseudoscalar case 
we would expect them to be also present in the vector-vector final state.

We will first review
the study the masses and mixing in the vector meson sector following 
the simple nonrelativistic 
 approach in
Ref\cite{Cohen:1979ge,Frank:1984bj}.
To obtain the eigenstates and eigenvalues in the vector meson system we
diagonalize
the mass matrix 
\ber
< q_a'\bar{q}_b',n'|M|q_a \bar{q}_b,n> & = &
\delta_{aa'} \delta_{bb'} \delta_{nn'}
(m_a +m_b +E_n) +\delta_{aa'}\delta_{bb'}\frac{B}{m_am_b}
{ \vec{s}_a \cdot \vec{s}_b} \psi_n(0)\psi_{n'}(0)\
\label{mm}
\eer
where $\vec{s}_{a,b}$ and $m_{a,b}$ are the quark spin operators and masses.
Here $n=0,1,2$ and the basis states for the isovector  mesons are chosen as
$\ket{N,I=1,I_3=1}=-\ket{ u \bar{d}}$,
$\ket{N,I=1,I_3=0}=\ket{ u \bar{u} - d \bar{d}}/\sqrt{2}$ and
$\ket{N,I=1,I_3=-1}=\ket{ d \bar{u}}$ where $I,I_3$ stand for the 
isospin quantum numbers.
In the above equation $E_n$ is the excitation energy of the $n^{th}$ 
radially excited state and $B$ is the strength of the hyperfine interaction. 

To begin with, we use the same harmonic
confining potential as well as the other parameters
used in
Ref\cite{Frank:1984bj}
to obtain the eigenstates and
eigenvalues for the mass matrix in Eqn.~\ref{mm}. The various
parameters used in the calculation are the constituent masses, $m_u=m_d=$ 
0.350 GeV, $m_s=$ 0.503 GeV,
the angular frequency,
$\omega= $0.365 GeV and $b=B/m_u^2=$0.09.
The eigenvalues and eigenstates for the $\rho$ system with a
harmonic potential are shown in
in Table.~\ref{rhoquadratic}.
\begin{table}[thb]
\caption{Eigenvalues and Eigenstates for the $\rho$ system-Harmonic potential }
\begin{center}
\begin{tabular}{|c|c|c|c|c|}
\hline
Harmonic  & $N_0$ & $N_1$ & $N_2$ \\
\hline
$\rho(0.768)$ &
0.990 &0.124& -0.066 \\
\hline
$\rho(1.545)$ &
0.108 &-0.973& 0.204 \\
\hline
$\rho(2.370)$ &
0.089 &-0.195& 0.977 \\
\hline
\end{tabular}
\end{center}
\label{rhoquadratic}
\end{table}
To see how this result
changes with a different confining potential we use a power law
potential $V(r)= \lambda r^n$ \cite{Quigg:1979vr} . We will use a linear and 
a quartic
confining potential
and compare the spectrum with that obtained
with a
harmonic oscillator potential. To fix the coefficient $\lambda$
we require that the energy eigenvalues of the
Schr\"odinger equation
are similar
in the least
square sense with the energy eigenvalues used in
Ref\cite{Frank:1984bj}. So for example, for the linear potential, we
demand that
$$F=\sum_n (E_n(harmonic)-E_n(linear))^2$$ is a minimum. This fixes the
constant
$\lambda$ in $V(r)=\lambda r$ and we obtain
the eigenvalues and eigenstates
in Table.~\ref{rholinear}.
\begin{table}[thb]
\caption{Eigenvalues and Eigenstates for the $\rho$ system-Linear potential }
\begin{center}
\begin{tabular}{|c|c|c|c|c|}
\hline
Linear  & $N_0$ & $N_1$ & $N_2$ \\
\hline
$\rho(0.775)$ &
0.992 &0.112& -0.053 \\
\hline
$\rho(1.515)$ &
0.104 &-0.986& 0.130 \\
\hline
$\rho(2.260)$ &
0.066 &-0.122& 0.990 \\
\hline
\end{tabular}
\end{center}
\label{rholinear}
\end{table}
We follow the same procedure
for the quartic potential and obtain
the eigenvalues and eigenstates
in Table.~\ref{rhoquartic}
\begin{table}[thb]
\caption{Eigenvalues and Eigenstates for the $\rho$ system- Quartic potential }
\begin{center}
\begin{tabular}{|c|c|c|c|c|}
\hline
Quartic  & $N_0$ & $N_1$ & $N_2$ \\
\hline
$\rho(0.759)$ &
0.988 &0.129& -0.077 \\
\hline
$\rho(1.567)$ &
0.103 &-0.955& 0.278 \\
\hline
$\rho(2.370)$ &
0.11 &-0.267& 0.957 \\
\hline
\end{tabular}
\end{center}
\label{rhoquartic}
\end{table}
We observe from Tables. (1-3) that the mass eigenstates and
eigenvalues of the $\rho$ system are not very sensitive to the
confining potential and the
radial mixing effects are  small.

To obtain the eigenstates and eigenvalues in the
$\omega- \phi$ system we
diagonalize
the mass matrix
\ber
< q_a'\bar{q}_b',n'|M|q_a \bar{q}_b,n> & = &
\delta_{aa'} \delta_{bb'} \delta_{nn'}
(m_a +m_b +E_n) +\delta_{aa'}\delta_{bb'}\frac{B}{m_am_b}
{\vec{s}_a \cdot \vec{s}_b} \psi_n(0)\psi_{n'}(0)
\nonumber\\
& + & \delta_{ab} \delta_{a'b'} \frac{A}{m_am_b}
\psi_n(0)\psi_{n^{\prime}}(0)\
\label{massmatrix}
\eer
This has a similar structure as the $\rho$ system but now we have the
additional annihilation  interaction
with strength $A$ that causes
flavor mixing.
\cite{Cohen:1979ge,georgi}.

Diagonalizing
the mass matrix in Eqn.~\ref{massmatrix}, with the basis states
 $\ket{N}=\ket{ u \bar{u} + d \bar{d}}/\sqrt{2}$ 
and $\ket{S}=\ket{s\bar{s}}$, 
 we obtain the eigenvalues and the 
eigenstates of the $\omega-\phi$ system.
We use the same value for
the hyperfine interaction as used for the $\rho$ system.
For the linear potential we obtain with $B=0.09m_u^2$ and $A=0.005m_u^2$
the eigenstates and eigenvalues in Table.~\ref{philinear}.
\begin{table}[thb]
\caption{Eigenvalues and Eigenstates for the $\omega-\phi$ system- Linear potential }
\begin{center}
\begin{tabular}{|c|c|c|c|c|c|c|c|}
\hline
Linear  & $N_0$ & $N_1$ & $N_2$ & $ S_0$ & $S_1$& $ S_2$\\
\hline
$\omega(0.782)$ &
0.991 &0.123& -0.058 & -0.014& 0.004& -0.002 \\
\hline
$\phi(1.05)$ &
0.012 &0.011& -0.004 & 0.997 &0.071 &-0.034 \\
\hline
$\omega(1.52)$ &
-0.113 & 0.982 & 0.144& -0.006& -0.034 & 0.004 \\
\hline
$\phi(1.66)$ &
0.007 &-0.030& -0.014 & 0.068 &-0.994 &-0.077 \\
\hline
\end{tabular}
\end{center}
\label{philinear}
\end{table}
For the harmonic potential we obtain with $B=0.09m_u^2$ and $A=0.015m_u^2$
the eigenstates and eigenvalues in Table.~\ref{phiquadratic}.
\begin{table}[thb]
\caption{Eigenvalues and Eigenstates for the $\omega-\phi $ system- Harmonic potential }
\begin{center}
\begin{tabular}{|c|c|c|c|c|c|c|c|}
\hline
Harmonic  & $N_0$ & $N_1$ & $N_2$ & $ S_0$ & $S_1$& $ S_2$\\
\hline
$\omega(0.783)$ &
0.984 &0.154& -0.081 & -0.033& 0.011& -0.007 \\
\hline
$\phi(1.05)$ &
0.026 &0.029& -0.011 & 0.994 &0.089 &-0.048 \\
\hline
$\omega(1.57)$ &
-0.126 & 0.948 & 0.256& -0.008& -0.139 & 0.010 \\
\hline
$\phi(1.68)$ &
0.025 &-0.12& -0.07 & 0.082 &-0.976 &-0.143 \\
\hline
\end{tabular}
\end{center}
\label{phiquadratic}
\end{table}
For the quartic potential we obtain with $B=0.09m_u^2$ and $A=0.023m_u^2$
the eigenstates and eigenvalues in Table.~\ref{phiquartic}.
\begin{table}[thb]
\caption{Eigenvalues and Eigenstates for the $\omega-\phi$ system- Quartic potential }
\begin{center}
\begin{tabular}{|c|c|c|c|c|c|c|c|}
\hline
Quartic  & $N_0$ & $N_1$ & $N_2$ & $ S_0$ & $S_1$& $ S_2$\\
\hline
$\omega(0.783)$ &
0.980 &0.163& -0.096 & -0.049& 0.012& -0.009 \\
\hline
$\phi(1.05)$ &
0.041 &0.034& -0.015 & 0.991 &0.100 &-0.060 \\
\hline
$\omega(1.58)$ &
-0.122 & 0.932 & 0.322& -0.010& -0.113 & 0.006 \\
\hline
$\phi(1.7)$ &
0.022 &-0.089& -0.067 & 0.086 &-0.968 &-0.207 \\
\hline
\end{tabular}
\end{center}
\label{phiquartic}
\end{table}
As in the $ \rho$ system we find the mixing to be insensitive to the
confining potential and  the effects of radial mixing
to be small. We also find, as expected,
a small value for the
annihilation term in the fits to the masses.

We now use these wavefunctions to predict the ratios 
in Eq. (1-4).
These decays are dominated by diagrams which satisfy the inactive
spectator approach \cite{Datta} and are treated with Eq.(7-8). Some
of the diagrams which violate this assumption; e.g.  penguin and
annihilation contributions, may not be as negligible here as in the
charmonium case treated in Ref\cite{Datta} for the decays to the ground
state configurations. But they are expected to have much smaller form
factors for radial excitations. Therefore it is reasonable to neglect them
for this preliminary investigation of the order of magnitude of these
ratios.
Note that
for the $ B_s \rightarrow \pi^0 \phi$ decay the QCD penguin 
is isospin forbidden, the
annihilation contribution is OZI forbidden and the electroweak penguin is
also described by Eq. \ref{had}.
 
We  obtain, using 
 factorization for the nonleptonic amplitude,
\ber
R_{\rho^+}  & = & \left|\frac{\bra{\rho^{+ 
\prime}}\bar{u}\gamma^{\mu}(1-\gamma_5)b\ket{\bar{B}^0}
\bra{\pi^-}\bar{d}\gamma_{\mu}(1-\gamma_5)u\ket{0}}
{\bra{\rho^{+ }}\bar{u}\gamma^{\mu}(1-\gamma_5)b\ket{\bar{B}^0}
\bra{\pi^-}\bar{d}\gamma_{\mu}(1-\gamma_5)u\ket{0}}\right |^2 \nonumber\\
& = & \left |\frac{A_0^{\prime \rho^+}}{A_0^{\rho^+}}\right 
|^2\frac{P_{\rho^{+ \prime}}^3}
{P_{\rho}^3}\
\eer
where $P$ is the magnitude of the three momentum of the final states and the 
form factor $A^0$ is defined through
\ber
\bra{V_f}A_{\mu}\ket{P_i} & = &
(M_i +M_f)A_1\left[\epsilon_{\mu}^* -\frac{\epsilon^*.q}{q^2}q_{\mu}\right]
-A_2 \frac{\epsilon^*.q}{M_i+M_f}\left[
(P_i+P_f)_{\mu} -\frac{M_i^2-M_f^2}{q^2}q_{\mu}\right] \nonumber\\
& + & 2M_f A_0 \frac{\epsilon^*.q}{q^2}q_{\mu}\
\eer
where $A_{\mu}$ is the axial vector current.
Similarly we obtain
\ber
R_{\rho^0}  & \approx & \left|\frac{\bra{\rho^{0 
\prime}}\bar{u}\gamma^{\mu}(1-\gamma_5)b\ket{\bar{B}^0}
\bra{\pi^-}\bar{d}\gamma_{\mu}(1-\gamma_5)u\ket{0}}
{\bra{\rho^{0 }}\bar{u}\gamma^{\mu}(1-\gamma_5)b\ket{\bar{B}^0}
\bra{\pi^-}\bar{d}\gamma_{\mu}(1-\gamma_5)u\ket{0}}\right |^2 \nonumber\\
& = & \left|\frac{A_0^{\prime \rho^0}}{A_0^{\rho^0}}\right 
|^2\frac{P_{\rho^{0 \prime}}^3}{P_{\rho^0}^3}\
\eer
\ber
R_{\omega}  & \approx & \left|\frac{\bra{\omega^{ 
\prime}}\bar{u}\gamma^{\mu}(1-\gamma_5)b\ket{\bar{B}^0}
\bra{\pi^-}\bar{d}\gamma_{\mu}(1-\gamma_5)u\ket{0}}
{\bra{\omega}\bar{u}\gamma^{\mu}(1-\gamma_5)b\ket{\bar{B}^0}
\bra{\pi^-}\bar{d}\gamma_{\mu}(1-\gamma_5)u\ket{0}}\right|^2 \nonumber\\
& = & \left|\frac{A_0^{\prime 
\omega}}{A_0^{\omega}}\right|^2\frac{P_{\omega^{ \prime}}^3}{P_{\omega}^3}\
\eer
Finally,
\ber
R_{\phi}  & = & \left|\frac{\bra{\phi{ 
\prime}}\bar{s}\gamma^{\mu}(1-\gamma_5)b\ket{\bar{B}_s}
\bra{\pi^0}\bar{u}\gamma_{\mu}(1-\gamma_5)u\ket{0}}
{\bra{\phi}\bar{s}\gamma^{\mu}(1-\gamma_5)b\ket{\bar{B}^0}
\bra{\pi^0}\bar{u}\gamma_{\mu}(1-\gamma_5)u\ket{0}}\right|^2 \nonumber\\
& = & \left|\frac{A_0^{\prime \phi}}{A_0^{\phi}}\right|^2\frac{P_{\phi{ 
\prime}}^3}
{P_{\phi}^3}\
\eer
To calculate the ratios we need the form factor $A_0$.
Note that  the wavefunctions for the various vector meson states
 are  not enough to calculate non leptonic
decay amplitudes. In particular,
with the factorization assumption for non leptonic decays, the calculations
of decay amplitudes require the  matrix elements of
current operators between the physical states. These matrix elements
can be expressed in terms of form factors and decay constants.
In this work
we use a constituent quark model(CQM)
model for the form factors  \cite{Aleksan:1995bh}
that incorporates
some relativistic features
and is relatively simple to adapt to the case of radially excited states.
In this model the form factor $A_0$ is given by
\bers
A_0 & = & Z[I_1-\frac{M_{-}}{M_{+}}I_2]\
\eers
where
\ber
Z & = & \frac{\sqrt{4M_iM_f} M_{+}}{M_{+}^2-q^2}\nonumber\\
I_1 & = & \int d^3p \phi_f^{*}(\vec p + \vec a) \phi_i(\vec p) \nonumber\\
I_2 & = & m_s\int d^3p \phi_f^{*}(\vec p + \vec a) \phi_i(\vec p)
[\frac{\vec p. \vec a}{\mu a^2} + \frac{1}{m_f}]\nonumber\\
M_{\pm} & = & M_f \pm M_i \nonumber\\
\vec a & = &2 m_s \vec{\beta} =2 \frac{m_s \tilde {q}}{M_{+}} \nonumber\\
\tilde{q}^2 & =& M_{+}^2 \frac{M_{-}^2-q^2}{M_{+}^2-q^2} \nonumber\\
\mu & = & \frac{m_im_f}{m_i+m_f} \
\label{model1}
\eer
and $\phi_f$ and $\phi_i$ represent the momentum space wave functions while
$\vec{\beta}$
is the velocity of the mesons in the equal velocity frame
( also called the Breit frame or the brick wall frame) and $m_{i,f}$
are the non spectator quark masses of the initial and final meson.
 The equal velocity frame in a convenient frame to calculate the
Lorentz invariant form factors where the
velocities, $\vec{\beta_i}$ and $\vec{\beta}_f$
of the mesons with masses $M_i$ and $M_f$
are equal in magnitude but opposite in direction. 
We use the momentum wavefunction $\phi_f$ obtained from spectroscopy in
section. (2)
while for $\phi_i$ we use the wave function
\ber
\phi_i& = & \phi_B=N_B e^{-p^2/p_F^2} \
\label{phiB}
\eer
where $p_F$ is the Fermi momentum of the B meson. In our calculations we
will take
$p_F=300$MeV. Note that in the analysis presented in the introduction we
have
neglected the  Fermi momentum of the b quark, since
$p_F/m_b$ is small.  

For transitions to higher resonant states,
we use the same quark masses as those used in the
transition of the $B$ meson to the lowest resonant state. This is
reasonable, as
the spectator quark still comes from the
$B$ meson and therefore has the same value for its mass
irrespective of whether the final state is in the lowest or
the first excited state. The values for the masses of the
non spectator masses are taken to be   the same
as those used  for spectroscopy. However for
the calculation of the velocity $\vec{\beta}$ and hence $\vec{a}$ defined
in Eqn.~\ref{model1} we use the physical mass of the higher resonant state.
\begin{table}[thb]
\caption{Ratios of branching ratios for different confining potentials}
\begin{center}
\begin{tabular}{|c|c|c|c|}
\hline
Ratio & Linear & Quadratic & Quartic  \\
\hline
$R_{\rho^+}$ &
$2.3$ & $2.0$ & $1.9$ \\
\hline
$R_{\rho^0}$ &
$2.3$ & $2.0$ & $1.9$ \\
\hline
$R_{\omega}$ &
$3.5$ & $2.5$ & $1.7$ \\
\hline
$R_{\phi}$ &
$6.7$ & $6.2$ & $5.2$ \\
\hline
\end{tabular}
\end{center}
\label{T1}
\end{table}
In  Table.~\ref{T1}
we give our predictions for the various ratios defined above.
We find that the transitions
to higher excited states can be comparable or enhanced relative to the
transitions to the ground state. From Table.~\ref{T1}
we see that
the ratios of branching ratios are slightly sensitive
to the confining potential
and
the ratios of branching ratios increase as we go from the quartic to
the linear potential. This is because the wavefunction
for the linear potential has a longer tail and hence more
high momentum components than the wavefunction
for the quadratic  and the quartic potentials. The wavefunction
for the quadratic potential, has in turn,
a longer tail and hence more
high momentum components than the wavefunction for the quartic potential.
Hence we would expect the hierarchy
$(A_{0})_{linear} >(A_{0})_{quadratic} >(A_{0})_{quartic} $
 and a similar one for the radially excited states
$(A_{0}^{\prime})_{linear} >(A_{0}^{\prime})_{quadratic} >
(A_{0}^{\prime})_{quartic} $
where $A_{0}$ and $A_0^{\prime}$ are the form factors for the 
transition of $B$ to the ground state and the
first radially excited state  of the meson $M$.
We see from from Table.~\ref{T1} that this
hierarchy
is maintained for the ratios of form factors and so we have
$(A_{0}^{\prime}/A_{0})_{linear} >(A_{0}^{\prime}/A_{0})_{quadratic} >
(A_{0}^{\prime}/A_{0})_{quartic}  $.
Note that the ratio of form factors also depend on the choice of the
Fermi momentum of the $B$ meson, as
a smaller(larger) Fermi momentum would
make the form factors more(less) sensitive to
the tail of the wavefunction of $M$, as well as mixing effects in
the wavefunction of the meson $M$. The effect of mixing between the
various radially excited states and the ground state is generally small.

We observe in Table.~\ref{T1}
that there can be a large enhancement for $R_{\phi}$. This decay is suppressed in the 
standard model.
One can get a rough estimate of the branching ratio for $B_s \to \phi \pi^0$
using factorization as
\bers
\frac{BR[B_s \to \phi \pi^0]}{BR[\bar{B}^0 \to \rho^+ \pi^-]} & \approx &
\frac{1}{2} \left|\frac{V_{ub}V_{us}^{*}(c_1+c_2/N_c) -V_{tb}V_{ts}^* 
3c_9/2}
{V_{ub}V_{ud}^{*}(c_2+c_1/N_c)}\right|
\approx 0.02 \
\eers
where we have neglected form factor  and phase space differences between
$B_s \to \phi \pi^0$ and
$\bar{B}^0 \to \rho^+ \pi^-$. The Wilson coefficients $c_i$ 
can be found in Ref\cite{H_eff} while
$V_{ub},V_{us},V_{ud},V_{tb}$ and $V_{ts}$ are the various CKM elements
\cite{rpp2000}. 
Using the measured
$BR[ B \to \rho^{+} \pi^-] \sim 28 \times 10^{-6}$ \cite{Jessop:2000bv}  we 
get
$BR[B_s \to \phi \pi^0]  \sim 6 \times 10^{-7}$. Hence the large enhancement
for $R_{\phi}$ indicates that
  $BR[B_s \to \phi^{\prime} \pi^0]$ can be around $ \sim 4 \times 10^{-6}$.

Note that in the $\rho(\omega)$ system there are two resonances,
$\rho(1450)[\omega(1420)]$
and $\rho(1700)[\omega[1650]$, which can be identified
a S-wave radial excitation(2S) and a
D wave orbital excitation in the quark model. However
recent studies of the decays of
these resonances show that it is possible that these states are mixtures of
$q \bar{q}$ and hybrid
states Ref\cite{rpp2000}. Hence the state $\rho(1450)[\omega(1420)]$
is interpreted as
a 2S state with a small mixture of a hybrid state. We do not take
into account such possible mixing with a hybrid state in our
calculation and the meson masses
for these excited states used in our calculation are the ones
we predict in section 2.  For the $\phi$ system
there is only state at $\phi(1680)$ which we interpret as a 2S state in the 
absence of mixing effects.

In conclusion
we have considered the weak decays of a B meson to final states
that are mixtures
of S-wave radially excited components. We calculated nonleptonic
decays of the type $B \to \rho' \pi/B \to \rho \pi$,
$B \to \omega' \pi/B \to \omega \pi$ and
$B \to \phi' \pi/B \to \phi \pi$
where $\rho'$, $\omega'$ and $\phi'$ are higher $\rho$, $\omega$
and $\phi$ resonances. We found that the transitions to the excited
states can be comparable or enhanced relative to transitions
to the ground state. It would, therefore, be
extremely interesting to test these
predictions.
We also studied the effect of radial mixing in the vector system
generated from hyperfine interaction
and the annihilation term; these turn out to be generally small.

\centerline{ {\bf  Acknowledgment}}
This work was  supported  by the
US-Israel Bi-National Science Foundation
and by the Natural  Sciences and Engineering Research  Council
of Canada.


\begin{thebibliography}{99}
\bibitem{Datta}
A.~Datta, H.~J.~Lipkin and P.~J.~O'Donnell,
arXiv:hep-ph/0111336. To appear in Phys. Lett. B.

\bibitem{Beneke}
M.~Beneke, G.~Buchalla, M.~Neubert and C.~T.~Sachrajda,
hep-ph/0104110.

\bibitem{Cohen:1979ge}
I.~Cohen and H.~J.~Lipkin,
Nucl.\ Phys.\ {\bf B151}, 16 (1979).


\bibitem{Frank:1984bj}
M.~Frank and P.~J.~O'Donnell,
Phys.\ Rev.\ D {\bf 29}, 921 (1984).

\bibitem{newpaper} 
A.~Datta, H.~J.~Lipkin and P.~J.~O'Donnell,
arXiv:hep-ph/0102070;
Alakabha Datta, Harry J. Lipkin and 
Patrick J. O'Donnell. Work in progress.


\bibitem{Quigg:1979vr}
C.~Quigg and J.~L.~Rosner,
Phys.\ Rept.\ {\bf 56}, 167 (1979).

\bibitem{georgi}
A.~De Rujula, H.~Georgi and S.~L.~Glashow,
Phys.\ Rev.\ D {\bf 12}, 147 (1975);
N.~Isgur,
Phys.\ Rev.\ D {\bf 12}, 3770 (1975);
H.~Fritzsch and J.~D.~Jackson,
Phys.\ Lett.\ B {\bf 66}, 365 (1977).

\bibitem{H_eff}
A.~J.~Buras,
arXiv:hep-ph/9806471.

\bibitem{Aleksan:1995bh}
R.~Aleksan, A.~Le Yaouanc, L.~Oliver, O.~Pene and J.~C.~Raynal,
Phys.\ Rev.\ D {\bf 51}, 6235 (1995)
[hep-ph/9408215].

\bibitem{rpp2000} 
D. E. Groom {\it {et al.}} 
The European Physical Journal {\bf {15}} (2000) 1.

\bibitem{Jessop:2000bv}
C.~P.~Jessop {\it et al.}  [CLEO Collaboration],
Phys.\ Rev.\ Lett.\ {\bf 85}, 2881 (2000)
[hep-ex/0006008].
\end{thebibliography}
\end{document}